\documentclass[conference]{IEEEtran}
\IEEEoverridecommandlockouts
\usepackage{cite}
\usepackage{amsmath,amssymb,amsfonts}
\usepackage{algorithmic}
\usepackage{graphicx}
\usepackage{textcomp}
\usepackage{xcolor}
\usepackage{epsfig}
\usepackage{amsfonts}
\usepackage{comment}
\usepackage[linesnumbered,ruled]{algorithm2e}
\usepackage{bm}
\usepackage{float}
\usepackage{enumitem}
\DeclareMathOperator*{\argmin}{argmin}
\newcommand*{\argminl}{\argmin\limits}

\newcommand{\norm}[1]{\left\lVert #1 \right\rVert}
\def\BibTeX{{\rm B\kern-.05em{\sc i\kern-.025em b}\kern-.08em
    T\kern-.1667em\lower.7ex\hbox{E}\kern-.125emX}}
\begin{document}

\title{Real time enhancement of operator's ergonomics in physical human - robot collaboration scenarios using a multi-stereo camera system}

\author{\IEEEauthorblockN{Gerasimos Arvanitis$^{\ast}$, Nikos Piperigkos$^{\dagger}$, Christos Anagnostopoulos$^{\dagger}$, Aris S. Lalos$^{\dagger}$, Konstantinos Moustakas$^{\ast}$}
\IEEEauthorblockA{\textit{$^{\ast}$Department of Electrical and Computer Engineering, University of Patras, Patra, Greece} \\ \textit{$^{\dagger}$Industrial Systems Institute, ATHENA Research Center, Patra, Greece} \\
arvanitis@ece.upatras.gr, piperigkos@ceid.upatras.gr, anagnostopoulos@isi.gr, lalos@isi.gr, moustakas@upatras.gr}
%\begin{comment}

%\end{comment}
}

\maketitle

\begin{abstract}
	In collaborative tasks where humans work alongside machines, the robot's movements and behaviour can have a significant impact on the operator's safety, health, and comfort. To address this issue, we present a multi-stereo camera system that continuously monitors the operator's posture while they work with the robot. This system uses a novel distributed fusion approach to assess the operator's posture in real-time and to help avoid uncomfortable or unsafe positions.
	The system adjusts the robot's movements and informs the operator of any incorrect or potentially harmful postures, reducing the risk of accidents, strain, and musculoskeletal disorders. The analysis is personalized, taking into account the unique anthropometric characteristics of each operator, to ensure optimal ergonomics. The results of our experiments show that the proposed approach leads to improved human body postures and offers a promising solution for enhancing the ergonomics of operators in collaborative tasks.
	%In collaborative tasks between humans and machines, the robot's "behaviour" and movement (e.g., trajectory, motion, pose, etc.) can potentially affect the operator's safety, health and comfort. In this paper, we propose a multi-stereo camera system, with a novel distributed fusion approach, that continuously monitors the pose of the operators, while they collaborate with the robot, and assists them to avoid uncomfortable and unsafe postures by i) appropriately adjusting the robot's movement and ii) informing the operator about a wrong or potential harmful posture, reducing the risk of accidents, strain and musculoskeletal disorders. The personalized analysis is a real-time risk assessment ergonomic methodology, based on the unique anthropometric characteristics of each operator. The experimental results show that the proposed approach improves the human body postures and offers a promising solution to enhance ergonomics of the operators. % To demonstrate the potential of the proposed framework, two different applications are considered.
\end{abstract}


\section{Introduction}
\label{sec:intro}
In manufacturing, when operators are physically interacting with a robot, trying to accomplish a collaborative task, their posture is inevitably influenced by the robot's movement and trajectory \cite{jbp:/content/journals/10.1075/is.18018.mak, 10.1007/978-3-031-11609-4_46}. Work-related Musculoskeletal Disorders (WMSD) is a major health problem in developed countries \cite{MAURICE201788} and there is a worldwide interest to reduce the conditions and risk factors \cite{9724640} that may cause this problem \cite{safety7040071}, decreasing in this way the substantial costs for therapies and negative impacts on the life quality of operators \cite{8794044}. To overcome this challenging situation, adapted systems are required, able to be adjusted according to the anthropometrics and special characteristics of each operator, improving their physical situation and ergonomics \cite{9444203}. In literature, there are a lot of works \cite{9503394} that have used depth sensors (Kinect) or low-cost RGB devices to compute the Rapid Upper Limb Assessment (RULA) \cite{Plantard2017ValidationOA, 10.3233/IDT-170292} score or to detect awkward postures \cite{MANGHISI2017481}. To achieve this, open-source landmark estimation libraries \cite{8765346, KIM2021103164} are used for calculating the body joint angles. Additionally, a lot of research has been done to improve physical ergonomics \cite{9707485} in human-robot collaboration \cite{CICCARELLI2022689, 2108.05971} by generating task assignments \cite{8278841, 8206107}, finding the optimal trade-off motions \cite{9601163}, providing wearable feedback \cite{8429071}, estimating the physical load required in a daily job \cite{9674843}, or providing an automatic assessment of ergonomics \cite{8620549}.

%In this work, we are based on the assumption that the collaborative tasks between robots and humans in an industrial environment are usually performed as a sequence of known and pre-defined actions since it is not safe to allow robots to improvise. A parameter that can affect the robot’s movement is the characteristics of the operator's body (i.e., height). This parameter does not change the main action of the robot (e.g., bring a tool) but can be used to select the best adjustment of the robot’s position and configuration in order to optimize the ergonomics assessment of the operator (e.g., bring the tool into the most comfortable place for the operator without the need of stressing). The adjustment happens only at the beginning of the robot's operation since, for security reasons, it is not appropriate to have a dynamic adjustment. The main contributions of this work can be summarized as follows:

In this study, we operate under the premise that collaborative tasks between robots and humans in industrial environments are typically comprised of pre-defined and known actions for the sake of safety. One factor that can impact the robot's movement is the operator's body characteristics, such as height. This factor does not alter the primary action of the robot (e.g., delivering a tool), but can be utilized to select the optimal adjustment of the robot's position and configuration for optimized ergonomics assessment of the operator (e.g., placing the tool in a comfortable location without causing strain).
This adjustment takes place only at the start of the robot's operation, as dynamic adjustments are deemed inappropriate for safety reasons. The adaptation of the robot's trajectory facilitates the improvement of the operator’s ergonomics and it is performed based on their unique anthropometric characteristics. The key contributions of this work can be summarized as follows:
%\vspace{-0.7em}
\begin{itemize}[leftmargin=*]
	\item A graph Laplacian-based scheme that provides a 3D landmark optimization, allowing for the best use of available captured information at any time, regardless of the camera source. 
	\item The development of a manufacturing simulator from scratch, emulating robotic movements and operator's actions in human-robot collaboration scenarios.
	\item A framework that increases the operators’ situational awareness regarding uncomfortable work postures that may be harmful in long-term exposure. This is achieved through a personalized anthropometric estimation component that provides a comprehensive ergonomic analysis of the operators' actions during their collaboration with the robot. 
	\item An offline tool, suitable for the ergonomics experts to analyse and evaluate the operator's action and help them to set personalized rehabilitation strategies or recommendations to avoid muscle injuries.
\end{itemize}
%\vspace{-0.7em}
\noindent The rest of this paper is organized as follows: In Section 2, we discuss in detail each step of the proposed method. Section 3  presents the experimental results and in Section 4 we draw the conclusions.
\section{General Architecture \& Methodology} 
\begin{figure}[htb]
	\centering
	\includegraphics[width=0.99\linewidth]{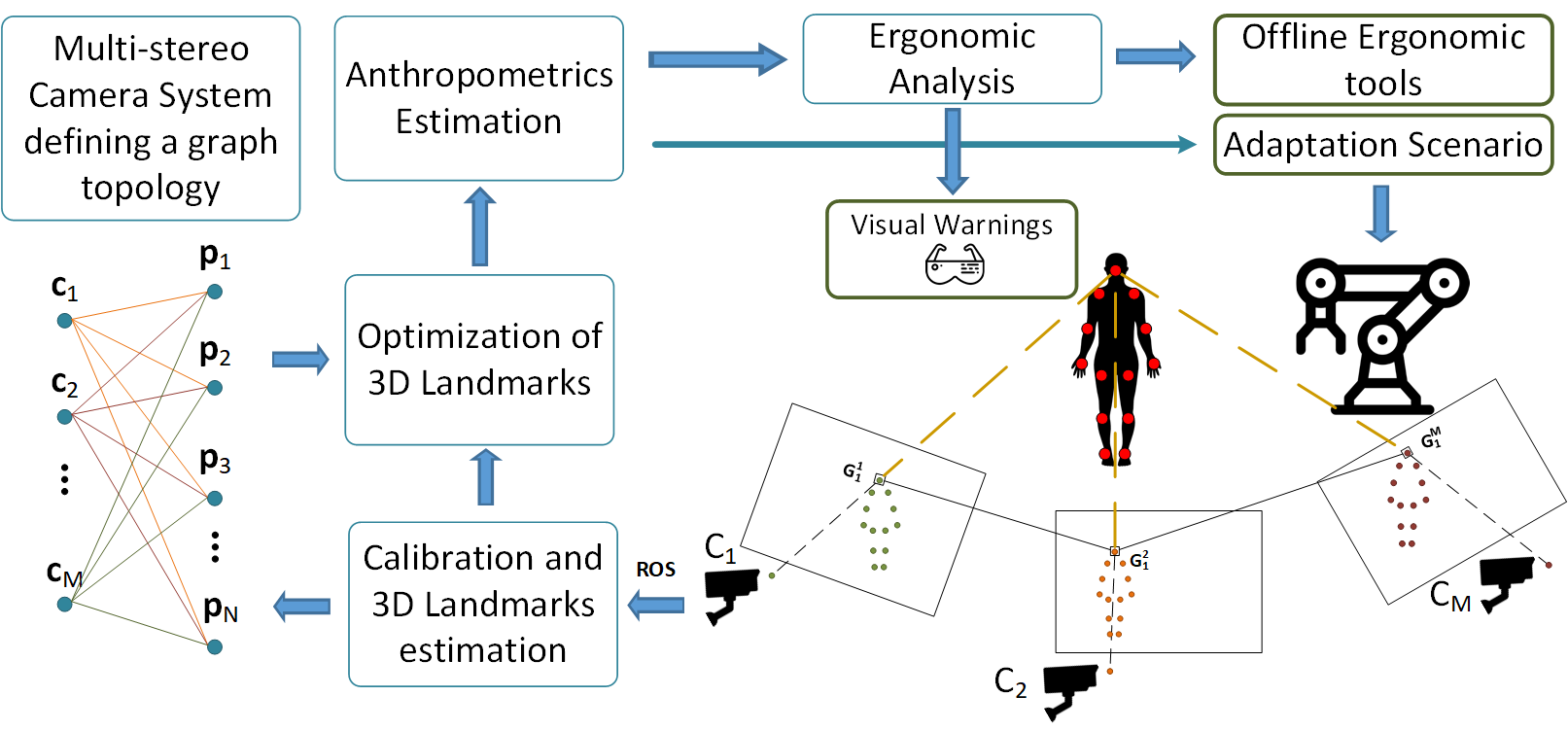}
	\vspace{-1.8em}
	\caption{Proposed concept architecture.}
	\label{fig:architecture}
\end{figure}

%	\vspace{-0.7em}
\label{framework}
%\subsection{General Architecture} 
The architecture of the proposed multi-stereo camera system is illustrated in Fig. \ref{fig:architecture}.  In our implementation, the system consists of three stereo cameras placed in strategic locations to provide continuous monitoring of the operator's pose from multiple angles, increasing the possibility of continuously having a good capture of the operator's pose (i.e., a view in front of the operator), at least from one of the cameras, while he/she is freely moving in different directions. 
%In our implementation, 3 stereo cameras are placed in such locations to cover the most possible visible area of the operators' working space, increasing the possibility of continuously having a good capture of the operator pose (i.e., a view in front of the operator), at least from one of the cameras, while he/she is freely moving in different directions and positions. 
%Each RGB camera of each stereo set is used to monitor the human’s actions. MediaPipe framework \cite{9929527} is running (locally to the Jetson device where the stereo camera is connected) to extract the posture 2D landmarks that are sent then to the PC server for the 3D landmark estimation based on a Direct Linear Transform (DTL) triangulation approach \cite{ABDELAZIZ2015103}, and next for the height estimation and to calculate the current physical ergonomic state (based on the RULA score\footnote{https://ergo-plus.com/rula-assessment-tool-guide}). 
We use each RGB camera of each stereo set to capture the operator's actions. To extract 2D posture landmarks, we utilize the MediaPipe framework \cite{9929527}, which runs locally on the Jetson device where the stereo camera is connected. The extracted landmarks are then sent to a PC server for 3D landmark estimation using a Direct Linear Transform (DTL) triangulation approach \cite{ABDELAZIZ2015103} and optimization. Next, we estimate the operator's height and calculate their current physical ergonomic state based on the RULA score\footnote{https://ergo-plus.com/rula-assessment-tool-guide}.
%For the seamless integration of the components and the algorithms, we chose the Robotic Operating System (ROS) as our middleware. Programs run on isolated nodes that can communicate using a publish-subscribe model. We have developed one ROS node for every camera used in the experiment. Each node captures a frame of the camera with a frequency of 10Hz, extracts and processes the landmarks, and finally publishes them to a relevant ROS topic. The decentralized approach of ROS enables us to connect the cameras to different physical machines, like the Jetson TX2 embedded device. A different node subscribes to the topics published by the cameras and extracts an optimal landmark set which will be used subsequently for the calculation of the posture of the operator. To notice here that Jetson devices and the server have to be connected to the same local network.
For seamless integration of the system components and algorithms, we utilize the Robotic Operating System (ROS) as our middleware. The system runs on isolated nodes that communicate using a publish-subscribe model, and we have developed a ROS node for each camera. Each node captures a frame of the camera at a frequency of 10Hz, extracts and processes the landmarks, and publishes them to a relevant ROS topic. The decentralized approach of ROS allows us to connect the cameras to different physical machines, such as the Jetson TX2 embedded device, and all components must be connected to the same local network.

Algorithm 1 summarizes the main steps of the proposed framework. More details will be presented in the next sections.
\begin{algorithm}%[H]
	\SetAlgoLined
	\tcc{A) Calibration steps (It is performed only once)}
	\For{$i = 1:M$}{Single camera $C_i$ calibration;}
	\For{$j = 1:L$}{Stereo camera $S_j$ calibration;}
	\KwData{Frames' sequency from the $M$ cameras}
	\KwResult{Improve operators' physical ergonomics}
	\tcc{B) 3D landmarks estimation}
	\For{$i = 1:M$}{	\tcc{It runs locally to the Jetson devices} 2D pose landmark estimation from $C_i$;}
	Send the 2D pose landmarks to the central server via ROS\;
	\For{$j = 1:L$}{3D pose landmark estimation from  $S_j$ via DLT}
	\tcc{C) 3D landmarks optimization}
%	\KwData{3D pose landmarks}
%	\KwResult{Optimized 3D pose landmarks}
	Follow the steps 1-5 of section II.B\;
	\tcc{D) Anthropometric and ergonomic analysis}
%	\KwData{Optimized 3D pose landmarks}
%	\KwResult{Operator's height estimation\; Real-time Rula score analysis}
Operator's height estimation\; Real-time RULA score analysis\;
	\label{algo}
		\caption{Algorithmic steps of the framework}
\end{algorithm}

\begin{figure*}[htb]
	\centering
	\includegraphics[width=0.28\linewidth]{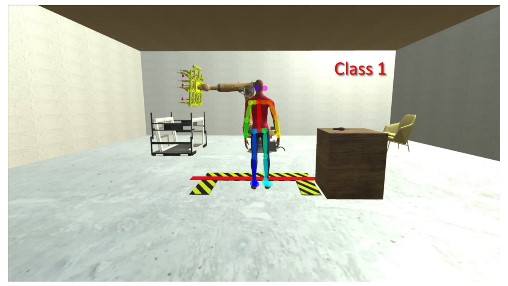}
	\includegraphics[width=0.28\linewidth]{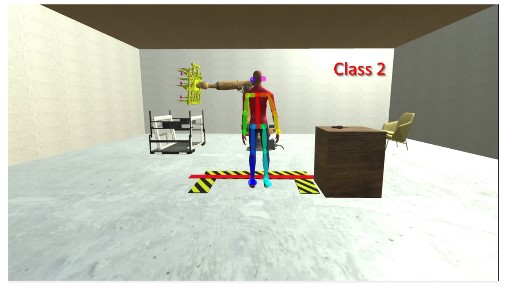}
	\includegraphics[width=0.28\linewidth]{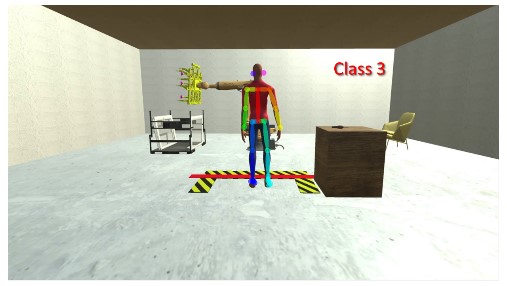}
		\vspace{-0.8em}
	\caption{Examples of the simulator in different configurations (height of the operator), corresponding to 3 different anthropometric classes.}
	\label{fig:classes}
\end{figure*}

\subsection{3D Landmarks Calculation} \label{3DLandmark}
%The steps that we have to follow for the calculation of the 3D landmarks are: 1) Calibration of each camera separately using a known and pre-defined checkerboard pattern, 2) Calibration of stereo cameras setup, 3) Use of the DLT algorithm to triangulate camera pixels to 3D coordinates.

%\subsubsection{Single Camera Calibration}
\noindent 1) \textbf{\textit{Single Camera Calibration}}. To calibrate each camera individually, we begin by using the same checkerboard pattern positioned in a location that is visible to all cameras at the same time. This step is crucial for performing stereo calibration later. We also ensure that all the used cameras have the same resolution analysis and capture images at the same frame-per-second (fps) ratio to ensure that all captured frames are synchronized \cite{9641587}. %To mention here that each camera is considered as a simple pinhole model having a coordinate system at the centre of the camera lens. Light rays from a distinct point in the camera’s view travel through the camera lens and converge into a single point in the camera. In the central projection camera model, this vector intersects an image plane at a distance $f$, which represents the focal length, along the $Z$ axis \cite{9641587}.

%\subsubsection{Stereo Calibration}
\noindent 2) \textbf{\textit{Stereo Calibration}}. Let's assume that each camera of the integrated system can be defined as $C_i \ (\forall \ i = 1, \cdots, 6)$ with $\textbf{c}_i \in \mathbb{R}^{3 \times 1}$ 3D coordinates. The single-camera calibration of camera $C_i$ returns the rotation matrix $\textbf{R}_i \in \mathbb{R}^{3 \times 3}$, the translation $\textbf{T}_i \in \mathbb{R}^{3 \times 1}$ matrix and the distortion coefficients. However, the $\textbf{R}$ and $\textbf{T}$ matrices alone are not enough to triangulate the 3D points. %So first, we have to define a world space origin point and orientation. 
For the stereo-calibration procedure, we need two sets of cameras. For the sake of simplicity, let's define these two cameras as $C_1$ and $C_2$. Notice here that the same approach can be used for the stereo calibration of all other cameras.
The rotation matrix $\textbf{R}$ and translation vector $\textbf{T}$ shows how to go from the $C_2$ coordinate system to the $C_1$. 
%The aforementioned approach does not contain any information about the world coordinate rotation matrix and translation vector, but only information related to the $C_2$ position and rotation concerning $C_1$. Nevertheless, we can overcome this problem, 
We also obtain world coordinates to $C_2$ rotation and translation, by calculating: $	\textbf{R}_2 = \textbf{R} \textbf{R}_1$, $\textbf{T}_2 = \textbf{R} \textbf{T}_1 + \textbf{T}$. Next, we choose the $C_1$ position as world space origin ($x=0, y=0, z=0$). Thus, the world origin to $C_1$ rotation is equal to the identity matrix and translation is a zeros vector. Then, the $\textbf{R}$ matrix and the $\textbf{T}$ vector that have been estimated from the previous stereo-calibration step becomes the rotation and translation from world origin to $C_2$. Basically, this means that the 3D triangulated points will be in a space with respect to the coordinate system of $C_1$'s lens. In other words, we simply overlap world coordinates with the coordinates of the $C_1$ camera. This means $\textbf{R}_1 = \text{eye}(3)$, $\textbf{T}_1 = \text{zeros}(3)$ and $\textbf{R}_2 = \textbf{R}$, $\textbf{T}_2 = \textbf{T}$. Therefore, all triangulated 3D points are measured from the $C_1$ camera position in the world.
%\begin{figure}[htb]
%	\centering
%	\includegraphics[width=0.48\linewidth]{images/calib1}
%    \includegraphics[width=0.48\linewidth]{images/calib2}
%	\vspace{-0.8em}
%	\caption{Example of the left and right camera calibration.}
%	\label{fig:calibration}
%\end{figure}

%\subsubsection{Direct Linear Transform}
%The 3D point triangulation from multiple camera via Direct Linear Transforms (DLT).
\noindent 3) \textbf{\textit{Direct Linear Transform}}. We assume the existence of a 3D point $\textbf{p}$ in real space with coordinates given as $\textbf{p} = [x, y, z]$. This point can be observed from both 2 cameras, which have pixel coordinates $\textbf{G}_i = [u_i, v_i, 1]$ for $C_i, \forall \ i = 1,2$. Using the camera projection matrix $\textbf{K}_1 = [\textbf{R}_1 \textbf{T}_1] \in \mathbb{R}^{3 \times 4}$, $\textbf{G}_1$ can be written as: ${\rm \textbf{G}}_{1} = a \textbf{K}_1$.
%\begin{equation}
%\small	\Vec{\rm \textbf{G}}_{1} = a \textbf{K}_1 \Vec{\textbf{p}}
%\end{equation}
In a triangulation problem, we do not know the coordinates of $\textbf{p}$. But we can determine the pixel coordinates and  the projection matrix through camera calibration. %Our task is then to determine the unknown values of $\textbf{p}$. 
Since $\textbf{G}_1$ and $\textbf{K}_1\textbf{p}$ are parallel vectors, the cross product of these should be zero. 
The $j$ row vector of $\textbf{K}_1$ can be written as $\textbf{k}_j \forall \ j = 1,...,4$. This gives us:
%\begin{equation}
%\small
\begin{multline}
%\footnotesize 
\begin{aligned}
\noindent	\begin{bmatrix}
u_1\\
v_1\\
1
\end{bmatrix}	 \times \begin{bmatrix}
{\textbf{k}}_1{\textbf{p}}\\
{\textbf{k}}_2{\textbf{p}}\\
{\textbf{k}}_3{\textbf{p}}
\end{bmatrix} =  
\begin{bmatrix}
v_1{\textbf{k}}_3{\textbf{p}}- {\textbf{k}}_2{\textbf{p}}\\
{\textbf{k}}_3{\textbf{p}}- u_1{\textbf{k}}_3{\textbf{p}}\\
u_1{\textbf{k}}_2{\textbf{p}}- v_1{\textbf{k}}_1{\textbf{p}} 
\end{bmatrix} = \\
\begin{bmatrix}
v_1{\textbf{k}}_3- {\textbf{k}}_2\\
{\textbf{k}}_3- u_1{\textbf{k}}_3\\
u_1{\textbf{k}}_2- v_1{\textbf{k}}_1 
\end{bmatrix} {\textbf{p}} = 
\begin{bmatrix}
0\\
0\\
0
\end{bmatrix}
\end{aligned}
\end{multline}
%\end{equation}
This gives us an equation of the form $\textbf{A}\textbf{x} = 0$. But the third row is a linear combination of the first two rows giving 2 systems of equations, which is not enough to solve the 3 unknowns in $\textbf{p}$. %, explaining why we can not estimate a 3D coordinate from a single camera. 
Since we have two cameras, we can extend the matrix to have more rows. In fact, we simply add on more rows for any number of views. This gives us the equation:
\begin{equation}
%\small
\begin{bmatrix}
v_1{\textbf{k}}_3- {\textbf{k}}_2\\
{\textbf{k}}_3- u_1{\textbf{k}}_3\\
u_1{\textbf{k}}_2- v_1{\textbf{k}}_1 \\
v_2{\textbf{k}}_3- {\textbf{k}}_2\\
{\textbf{k}}_3- u_2{\textbf{k}}_3\\
u_2{\textbf{k}}_2- v_2{\textbf{k}}_1 
\end{bmatrix} {\textbf{p}} = 0
\end{equation}
DLT is a method for calculating a matrix equation of the form $\textbf{A}\textbf{p} = 0$. In the real world, there can be some noise, so we write the equation as $\textbf{A}\textbf{p} = \textbf{n}$, and we solve for $\textbf{p}$ such that $\textbf{n}$ is minimized. The first step is to determine the SVD decomposition of $\textbf{A}$.
\begin{equation}
%\small	
\textbf{A} {\textbf{p}}   = \textbf{U}\textbf{S}\textbf{V}^{T} {\textbf{p}}
\end{equation}
Our goal is to minimize $\textbf{n}$ for some $\textbf{p}$. This can be done by taking the dot product:
\begin{equation}
%\small	
{\textbf{n}}^{T} {\textbf{n}} = ({\textbf{p}}^{T} \textbf{V} \textbf{S} \textbf{U}^{T}) \cdot ( \textbf{U} \textbf{S} \textbf{V}^{T} {\textbf{p}}) = {\textbf{p}}^{T} \textbf{V} \textbf{S}^{2} \textbf{V}^{T} {\textbf{p}}
\end{equation}
$\textbf{U}$ and $\textbf{V}$ are orthonormal matrices and $\textbf{S}$ is a diagonal matrix. The entries on the diagonal of $\textbf{S}$ are decreasing, so that the last entry on the diagonal is the minimum value. Since our goal was to minimize $\textbf{n}^{T}\textbf{n}$, this tells us that it is equivalent to choosing the smallest value of $\textbf{S}^2$ by selecting the corresponding column vector of  $\textbf{V}^{T}$ as $\textbf{p}$.%These are guaranteed by the SVD decomposition. Exploiting the property that V is an orthonormal matrix, if we simply select $\textbf{p}$ to be one of the column vectors of $V^T$:
%\begin{equation}
%\small	\Vec{u}^{T}_i V S^{2} V^{T} \Vec{u}_i = s_i^2
%\end{equation}
%In the above equation, we have written the i’th entry on the diagonal of S as si. %Since our goal was to minimize $w^{T}w$, this tells us that it is equivalent to choosing the smallest value of $S^2$ by selecting the corresponding vi column vector of VT as $\textbf{p}$. In other words, the minimum value is obtained if we choose the last column vector of VT as $\textbf{p}$. Thus we have solved the $\textbf{A}\textbf{p} = \textbf{n}$ equation in the presence of noise. If there is no noise, this SVD method will still work \cite{9157707}.
%\vspace*{-1.8mm}
\subsection{3D Landmark Optimization Algorithm} \label{3DLandmark_opt}
%\vspace*{-0.8mm}
As a result of varying camera viewpoints and operator movements, we expect that landmarks detected from different cameras will not be the same and may differ in accuracy. Therefore, it is necessary to design a fusion approach that integrates the multi-camera system to provide the best possible landmarks configuration from all the involved cameras.
\begin{figure}[htb]
	\centering
	\includegraphics[width=0.65\linewidth]{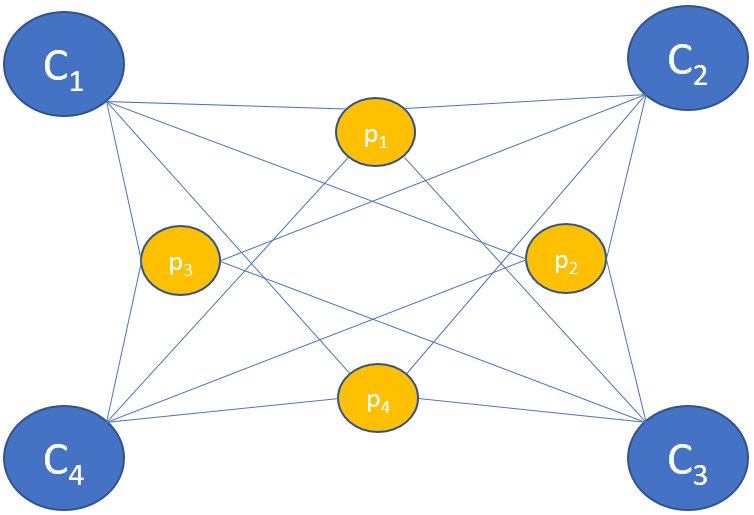}
	%\vspace{-0.8em}
	\caption{Indicative example of graph topology for cameras and landmarks.}
	\label{fig:graph}
\end{figure}
For this task, we create an undirected graph of $N+M$ points $\hat{\textbf{p}} = [\textbf{p} \ \textbf{c}]^{T} \in \mathbb{R}^{(N+M)\times 3}$, consisting of the $N$ detected landmarks as edges $\textbf{p}$, and the $M$ cameras as nodes $\textbf{c}$. Fig \ref{fig:graph} presents a simplified example of a graph with $M=4$ and $N=4$. In order to take advantage of this graph topology formulation, we apply the Graph Laplacian Processing (GLP) technique \cite{9334558} for re-estimating landmarks in an optimal manner, following the next steps: 

1) Encode the spatial relationship of  cameras $C$ and landmarks $\textbf{p}$ via GLP.
 
2) Construct the binary Laplacian matrix $\boldsymbol{L} \in \mathbb{R}^{(N+M)\times(N+M)}$ of connectivity graph, indicating which landmark is related to which camera. To mention here that there are not connections between two cameras and between two landmarks. 

3) Use as anchor points $\textbf{a} = [\bar{\textbf{p}} \ \textbf{c}]^{T} \in \mathbb{R}^{(N+M)\times3}$ the known location of cameras $\textbf{c}$ and the average position of landmarks $\bar{\textbf{p}}^i=\sum_{i=1}^M\textbf{p}^i \in \mathbb{R}^{N\times3}$ as estimated by each one of the cameras. 

4) Relative measurements and anchors are used to formulate differential coordinates $\boldsymbol{\delta} \in \mathbb{R}^{(N + M) \times 3}$ \cite{SorkineHornung2005LaplacianMP, 9733486}, where the $i^{th}$ row of $\boldsymbol{\delta}_i$ is equal to:
\begin{multline}
\begin{aligned}
\noindent 
\boldsymbol{\delta}_i = [ \boldsymbol{\delta}_{xi} \ \boldsymbol{\delta}_{yi} \ \boldsymbol{\delta}_{zi}] =  \hat{\textbf{p}}_i - \frac{1}{|\Psi_{i}|}\sum_{j\in \Psi_{i}}\hat{\textbf{p}}_j
=  \boldsymbol{L}\hat{\textbf{p}}_i, \\ \ \ \forall \ i = 1, ..., N + M
\label{delta}
%\end{equation}
\end{aligned}
\end{multline}
where $\boldsymbol{l_i}$ is the $i$-row of Laplacian matrix, and $|\Psi_{i}|$ is the number of immediate neighbors of $\hat{\textbf{p}}_i$. 

5) The optimized positions of point $\hat{\textbf{p}}$ are given by the minimization of the following cost function:
%\begin{comment}
\begin{equation}
\label{ext_lapl1}
%\small 
\argminl_{\hat{\textbf{p}}}{\norm{\boldsymbol{\tilde{L}} \hat{\textbf{p}} -\boldsymbol{b}}_2^2}
\end{equation}
where $\boldsymbol{\tilde{L}} = 
[\boldsymbol{L} \
\mathbb{I}_{N+M}]^T \in \mathbb{R}^{2(N+M)\times(N+M)}$, $\mathbb{I}_{N+M}$ is the identity matrix and $\boldsymbol{b} = [\boldsymbol{\delta}  \ \textbf{a}]^T \in \mathbb{R}^{2(N+M)\times3}$.
Finally, the re-estimated positions of landmarks are derived from linear least-squares minimization:
\begin{equation}
%\small 
\hat{\textbf{p}} =  \left(\boldsymbol{L^TL}\right)^{-1}\boldsymbol{L^Tb}
\label{lapl_function}
\end{equation}
To note that during the simulation horizon the topology of landmarks and cameras remains unchanged. As such, the term $\left(\boldsymbol{L^TL}\right)^{-1}\boldsymbol{L^T}$ is constant and can be computed only once at the start of simulation, in order to be used for the successive frames. In that way, we reduce the computational time required for solving \eqref{lapl_function} since the costly operation of matrix inversion is performed only at the beginning of the process.

%contains at its first half the differential coordinates among the nodes of graph and the noisy locations (of $x$ coordinate) of landmarks and cameras at its second half. %Note that the anchor positions of landmarks is the average among the estimations from each one of the involved cameras. 
%\end{comment}
%In practice, we are interested only for landmarks' part of state vector $\boldsymbol{x}$. The same procedure is followed for estimating $y$ and $z$ coordinates, by employing the corresponding differential coordinates and anchor points at GLP.
%Following this methodology, we end up with a camera-and-landmarks configuration very close or even better than the optimal one. 

\section{Experimental Analysis and Results}
%\vspace*{-1.2mm}
\subsection{Experimental Setup and Implementations}	
%\vspace*{-0.99mm}
%The experiments were carried out using an Intel Core i7-4710 CPU @ 3.60GHz PC with 16 GB of RAM.
%\subsection{Devices and equipment (jetson, zed, etc) that are used}
%photos of the implementation
The operating system of the server is UBUNTU 20.04 which is compatible with ROS 1. All the algorithms are written in Python 3.10. There are no special requirements regarding the computational power of the server since the implementation can easily run to low-cost devices (e.g., Jetson).  
%\subsection{Applications} \label{applications}
%\subsection{Operators Height Estimation} \label{app1}
%Two different approaches can be used, for the estimation of the operator’s height. %: i) Geodesic distance between nose to ankles of the operator + 15 cm, ii) Geodesic distance between the right index finger to the left index finger. The experiments show that the second approach is less prone to errors and this is the one that is finally used during the pilot.

%\noindent 
1) \textbf{\textit{Operator's Height Estimation}}. For this use case two steps are followed: i) Estimation of the operator's height; ii) Adaptation of the robot's position. % based on the operator's height classification.
Once the operator's height has been estimated, the robot's control system receives the information and configures the robot's movement parameters according to the selected scenario. This adaptation ensures that the robot's movements are ergonomically comfortable for the interacting user. 3 different classes $c$ are used, based on the operators' height ($c_1 < 1.68, 1.68 \leq c_2 \leq 1.82, c_3>1.82$) (Fig. \ref{fig:classes}). Each class leads to an adaptable robot response, ensuring that the user can work in an optimal ergonomic state. %The algorithmic framework identifies the operators’ class, based on their height, and the corresponding selected scenario configures the cooperating robot’s movement parameters and relative trajectories to adapt to the environment in a way that is ergonomically most comfortable for the interacting user. 
%Based on the anthropometric information a decision-making module finds the right set of actions in order to minimize strain and ergonomic risk factors. Finally, the robot receives the control information that the wanted set of actions is carried out. A total of 3 different classes are used based on the operators’ height, leading to 3 adaptable robot responses, allowing the human to work in an optimal ergonomics state.
%--The algorithmic framework identifies the operators' class, based on their height, and the corresponding selected scenario configures the cooperating robot's movement parameters to adapt to the environment in a way that is ergonomically most comfortable for the interacting user. We assume the existence of 3 different classes based on the operator’s height. More specifically, class 1 consists of operators with a height $<$ 175 cm, in class 2 the operators’ height is between 175 and 185 cm, and finally in class 3 the operators are taller than 185 cm.
%\\
%We assume the existence of 3 different classes based on the operator’s height.  More specifically: i) Class 1 - Operators with height $<$ 175 cm, ii) Class 2 - Operators’ height is between 175 and 185 cm, iii) Class 3 - Operators are taller than 185 cm. 
%\subsection{Real Time RULA Score Estimation and Warnings} \label{app2}

\begin{figure}[htb]
	\centering
	\includegraphics[width=0.99\linewidth]{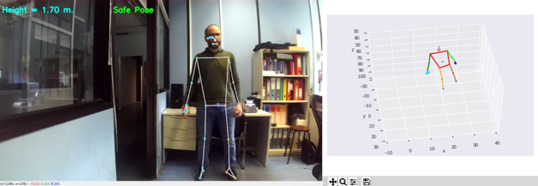}
	\includegraphics[width=0.99\linewidth]{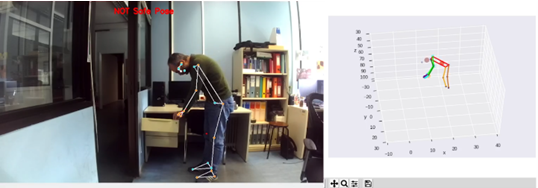}
	\vspace{-1.8em}
	\caption{ Safe and wrong pose identification.}
	\label{fig:safe_pose}
\end{figure}

%\noindent 
2) \textbf{\textit{Real-Time RULA Estimation}}. We perform a real-time joint angle estimation to calculate the RULA score (ranging from 1 to 7, used by ergonomists to evaluate work activities involving upper-body motion \cite{MCATAMNEY199391}). Based on this value, the appropriate warning messages are sent to the operators, informing them if their working pose is ergonomically correct or not (Fig. \ref{fig:safe_pose}), so as to correct their posture in real-time. 
%Simulations in a virtual environment were also used for further ergonomic analysis in order to optimize the workstation (Fig. \ref{fig:classes}), so as to obtain a trajectory of the robot, in a real environment, as much as possible adapted to the human and additionally to be used as a validation framework since the ground truth information of the 3D landmarks is known.
%\vspace*{-2.8mm}
%\vspace*{-2.8mm}

\begin{figure}[htb]
	\centering
	\includegraphics[width=0.9\linewidth]{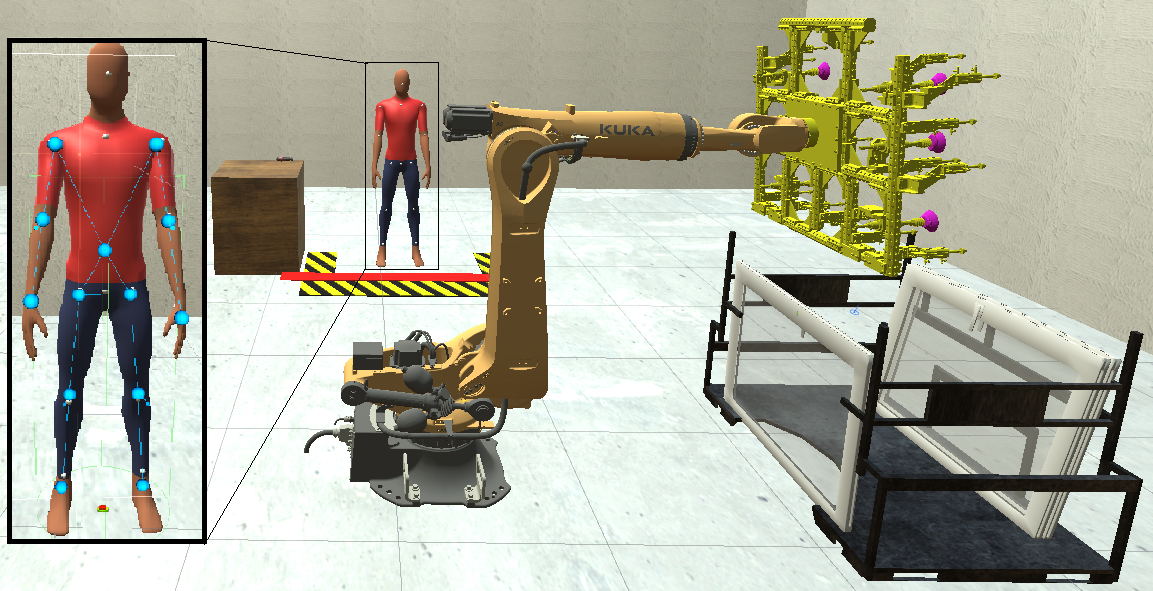}
	\vspace{-0.8em}
	\caption{\small Screenshot from the environment of the simulator.}
	\label{fig:simulator}
\end{figure}

\begin{figure}[htb]
	\centering	\includegraphics[width=0.99\linewidth]{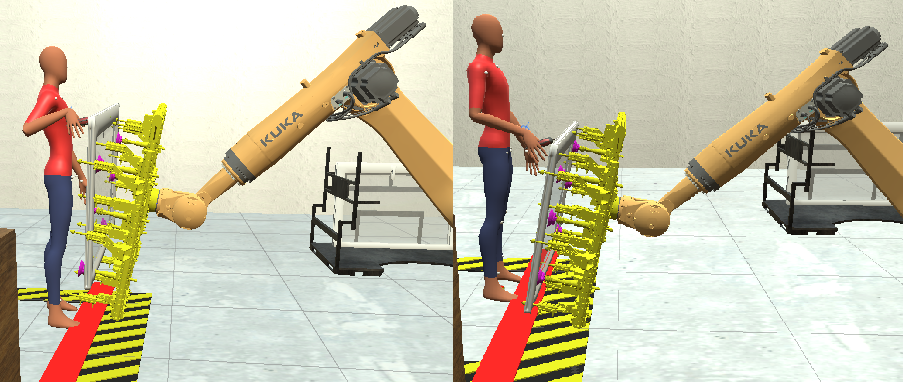}
	\vspace{-1.8em}
	\caption{Screenshot from the environment of the simulator before and after robot's adaptation.}
	\label{fig:adaptation}
\end{figure}

%\noindent 
3) \textbf{\textit{Simulator}}. To evaluate the effectiveness of our framework, we have developed an in-house simulation framework from scratch (Fig. \ref{fig:simulator}). The simulator emulates a robotic movement and the corresponding operator's actions in a digital workspace created in Unity3D \cite{Pavlou2021}. Fig. \ref{fig:adaptation} presents an example of a user's action before and after the robot's adaptation. 3 stereo cameras are involved, located in 3 different areas of the working environment, monitoring the operator's actions while performs a collaborative tasks with a KUKA robot. The primary objective of the manufacturing simulator is to create a digital twin, which provides a safe environment for evaluating the developed framework and algorithms. Additionally, the simulator can be also used for further ergonomic analysis in order to optimize the workstation of a real industrial environment, so as to obtain a trajectory of the robot as much as possible adapted to the human. Furthermore, the simulator serves as a validation framework since the ground truth information of the 3D landmarks' locations is not known in the real case.

\begin{figure}[htb]
	\centering
	\includegraphics[width=0.98\linewidth]{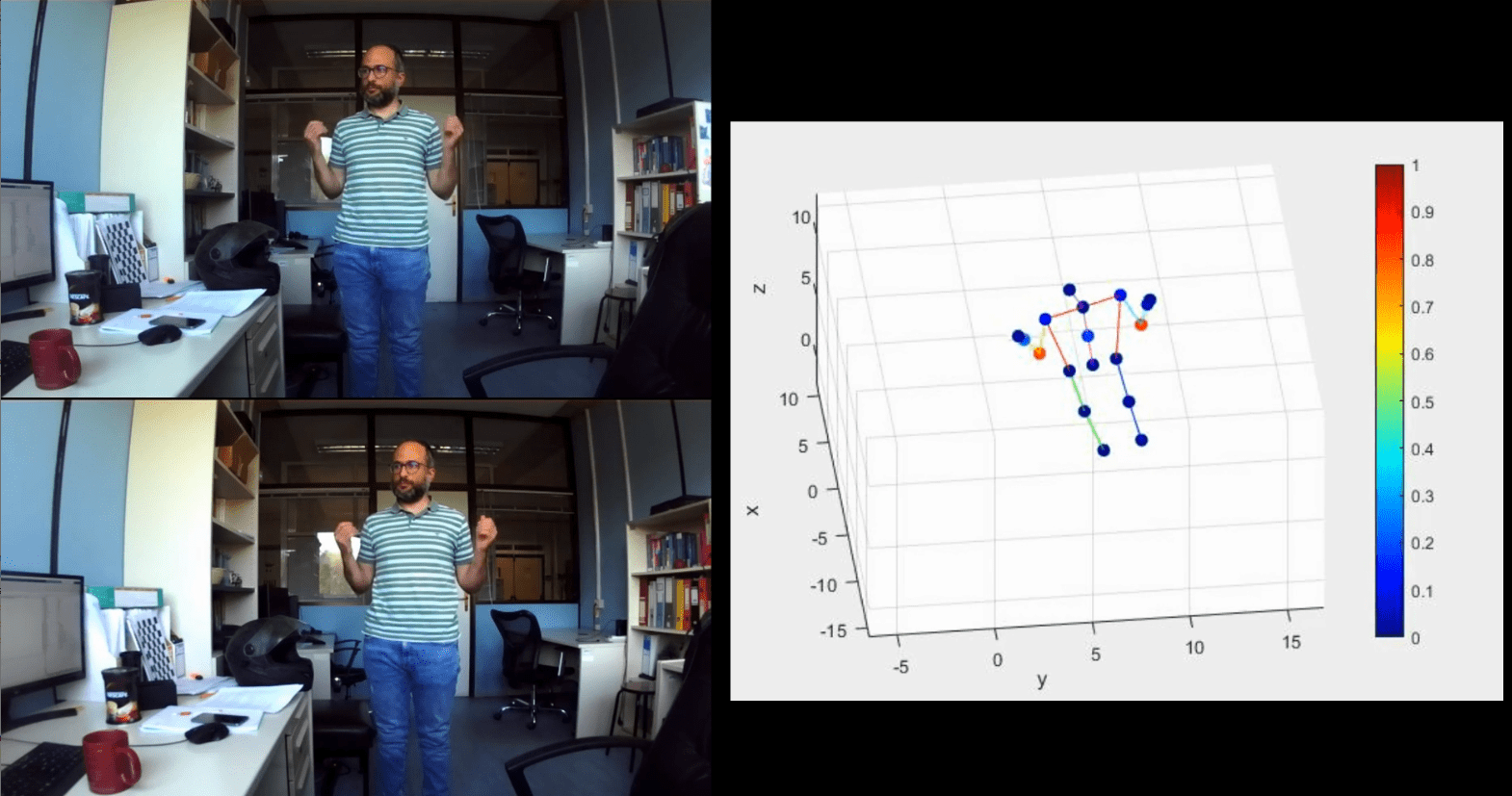} \\
	\includegraphics[width=0.98\linewidth]{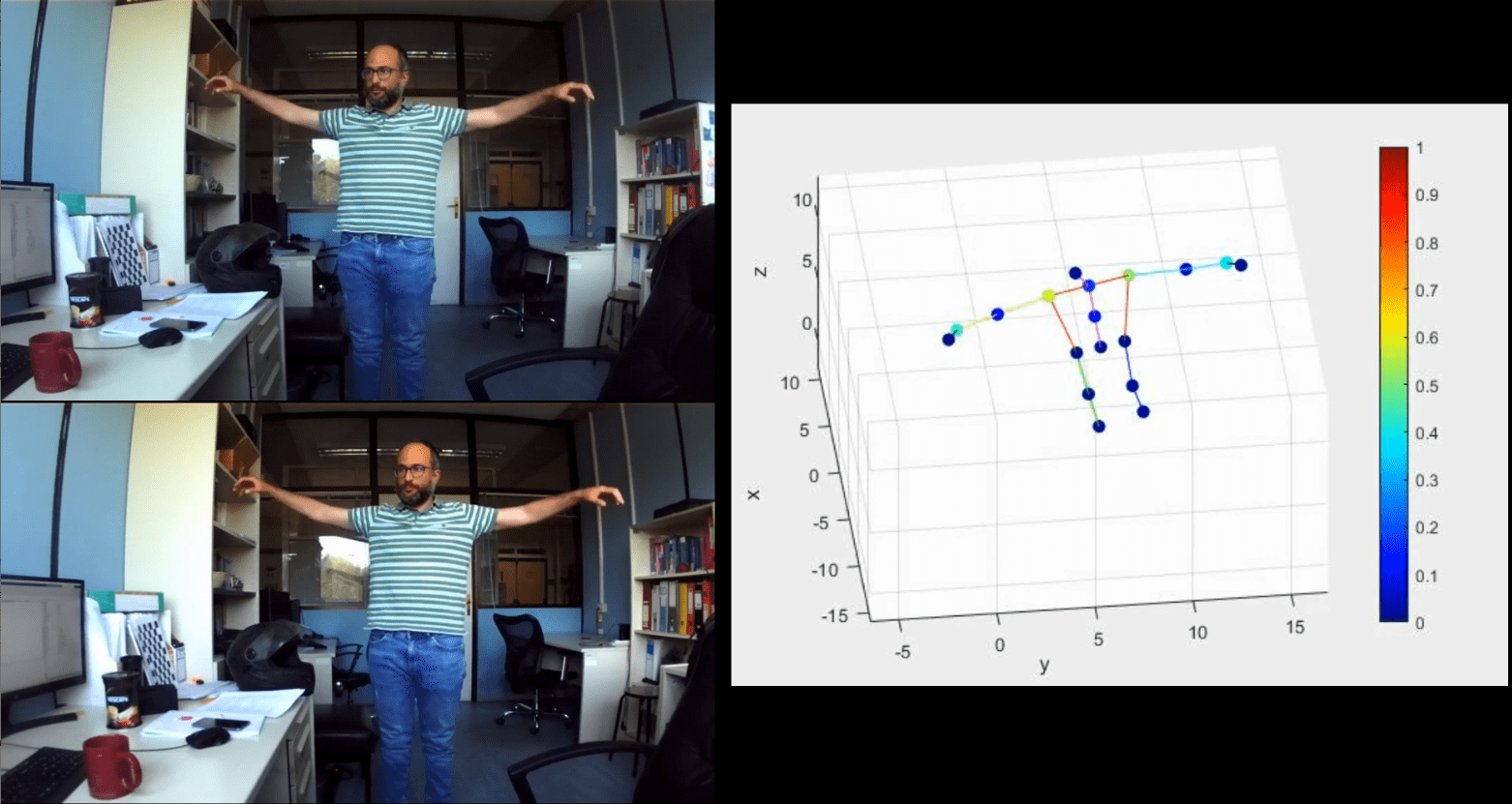} %\\
		\vspace{-0.4em}
	\caption{ Ergonomics tool for the evaluation of the joint angles stress.}
	\label{fig:rula_examples}
\end{figure}

 4) \textbf{\textit{Ergonomics Evaluation Tool}}. The 3D landmarks are stored and are accessible offline (i.e., 3D view, pause, zoom in/out) by ergonomic experts that can evaluate the poses during a specific action, and suggest how it could be improved. Fig. \ref{fig:rula_examples} illustrates an example showing how the joint angles are strained during an action. A heatmap visualizes the level of strength (deep blue is the lowest, deep red is the highest).
%Figure 23 and Figure 24 present screenshots of the integrated system that captures in real time the operator’s pose using two stereo-calibrated RGB cameras, estimate the 2D landmarks of the operator, then calculates the corresponding 3D landmarks and uses these landmarks to estimate and visualize the joint angles and finally to find the RULA score, per frame or per a sequence of 60 frames (6 seconds of action). More specifically, Figure 23 illustrates an example of a safe pose identification while Figure 24 illustrates a case of detection of a wrong pose that can affect the operator's wellbeing.

%\subsection{Simulations in a Digital Twin Environment for Testing and Evaluation} \label{sim}

%-- Screen shots of the Unity environment showing how we create operators in different heights, how we set the ground truth 3D landmarks, how we set up the cameras (position in the world coordinates and rotation, translation matrices between them) --- 
%\vspace*{-4.8mm}
\begin{table*}[htp]
	\caption{ RMSE of landmarks using multicamera system and GLP.} 
	\centering
	\vspace{-0.8em}
	\begin{tabular}{|c|c|c|c|c|c|c|c|c|c|c|c|c|}
		\hline
		\centering
		& \normalsize $\textbf{p}_1$ & \normalsize $\textbf{p}_2$ & \normalsize $\textbf{p}_3$ & \normalsize $\textbf{p}_4$ & \normalsize $\textbf{p}_5$ & \normalsize $\textbf{p}_6$ & \normalsize $\textbf{p}_7$ & \normalsize $\textbf{p}_8$ & \normalsize $\textbf{p}_9$ & \normalsize $\textbf{p}_{10}$ & \normalsize $\textbf{p}_{11}$ & \normalsize $\textbf{p}_{12}$ \\ \hline
		\normalsize $S_1$ & \normalsize  0.34 & \normalsize 0.38 & \normalsize 0.19 & \normalsize 0.24 & \normalsize 0.15 & \normalsize 0.18 & \normalsize 0.24 & \normalsize 0.27 & \normalsize 0.27 & \normalsize 0.27 & \normalsize 0.22 & \normalsize 0.35\\ \hline
		\normalsize $S_2$ & \normalsize 0.36 &  \normalsize 0.40 & \normalsize 0.19 &  \normalsize 0.28 & \normalsize 0.15 & \normalsize 0.21 & \normalsize 0.27 & \normalsize 0.29 & \normalsize 0.28 & \normalsize 0.32 & \normalsize 0.24 & \normalsize 0.38 \\ \hline
		\normalsize $S_3$ & \normalsize 0.56 & \normalsize 0.57 & \normalsize 0.34 & \normalsize 0.35 & \normalsize 0.32 & \normalsize 0.30 & \normalsize 0.41 & \normalsize 0.42 & \normalsize 0.43 & \normalsize 0.40 & \normalsize 0.31 & \normalsize 0.43 \\ \hline
		\normalsize fusion & \normalsize \textbf{0.38} & \normalsize \textbf{0.39} & \normalsize \textbf{0.20} & \normalsize \textbf{0.24} & \normalsize \textbf{0.16} & \normalsize \textbf{0.16} & \normalsize \textbf{0.27} &  \normalsize \textbf{0.27} &  \normalsize \textbf{0.30} & \normalsize \textbf{0.29} & \normalsize \textbf{0.21} & \normalsize \textbf{0.35} 
		\\ \hline
	\end{tabular}
	\label{rmse}
\end{table*}

\subsection{Experimental Results}
\begin{figure}%[htb]
	\centering
	\includegraphics[width=0.9\linewidth]{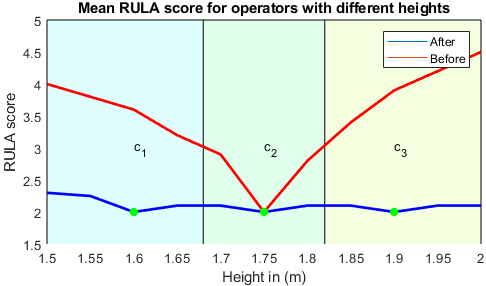}
	\vspace{-0.8em}
	\caption{ RULA score for operators with different heights.}
	\label{fig:example1}
\end{figure}

\begin{figure}[H]
	\centering
	\includegraphics[width=0.99\linewidth]{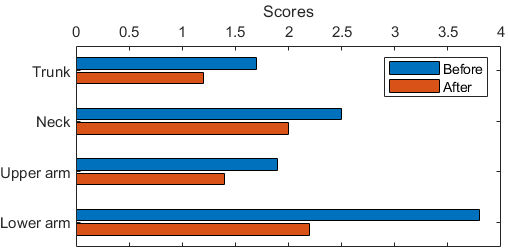}
    \vspace{-2.1em}
	\caption{ Mean score per different body areas.}
	\label{fig:example2}
\end{figure}

\begin{figure}[H]
	\centering
	\includegraphics[width=0.99\linewidth]{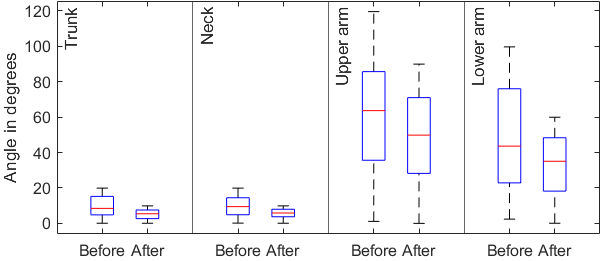}
	\vspace{-2.1em}
	\caption{ Distribution of angles per joint.}
	\label{fig:example3}
\end{figure}

Table \ref{rmse} summarizes the main outcomes (RMSE values of the 12 landmarks $\textbf{p}$, per each stereo camera $S_i \ \forall \ i \in (1-3)$) from a testing scenario running in the simulator. As we can see, in most of the cases, the proposed GF-based fusion approach achieved results very close to the optimal, while in two cases ($\textbf{p}_6$ and $\textbf{p}_{11}$), landmarks were detected slightly better than the optimal camera. To mention here that the evaluation of the fusion algorithm's performance can be achieved only in a simulated environment since in a real-case scenario the ground truth locations of the landmarks, as well as the accuracy of the cameras, can not be known. 
Fig \ref{fig:example1} shows a plot with the mean RULA score of different users with different heights (1.5-2 m.) that perform a very specific task before and after the robot's adjustment. As we can see, only operators with 1.75 m. height has a low RULA score before the adaptation, while it is improved for all of the users after the robot's adaptation.
Fig \ref{fig:example2} shows the mean score per different body areas for all operators. The improvement is more apparent for the cases of "Neck" and "Lower arms".
Fig \ref{fig:example3} shows how the distribution of the angles per each joint decreases after the robot's adaptation.
%\vspace{-2.2em}

%\vspace*{-3.8mm}
\section{Conclusions \& Future Work}
%\vspace*{-0.8mm}
%In this work, we presented a multi-stereo camera pipeline that provides an optimized way for the estimation of 3D landmarks that are used to improve the operator's ergonomics: 1) by changing the moving parameters of the collaborative machine (based on the unique operator's anthropometrics) and 2) by changing the motion and action of the operators (via on-line visual warnings or off-line expert's recommendations based on the RULA score estimation). A simulator is also developed to facilitate the evaluation of the integrated solution. Future plans include the implementation of the solution in a real industrial environment utilizing a survey for the evaluation of the method by real users' responses and evaluation in more complex collaborative actions.

In this work, we introduced a multi-stereo camera pipeline aimed at enhancing the operator's physical ergonomics by optimizing the estimation of 3D landmarks.  This is achieved in two ways: 1) by adjusting the movement parameters of the collaborative machine to the unique anthropometrics of the operator and 2) by modifying the operator's motion and actions through real-time visual warnings or expert recommendations, based on RULA score estimation. To facilitate the evaluation of our solution, we also developed a simulator. Future plans include implementation in a real industrial setting and a survey to gauge the effectiveness of the method through real user feedback and additionally the evaluation in more complex collaborative tasks.

% References should be produced using the bibtex program from suitable
% BiBTeX files (here: strings, refs, manuals). The IEEEbib.bst bibliography
% style file from IEEE produces unsorted bibliography list.
% -------------------------------------------------------------------------
%\newpage

%\footnotesize	
\bibliographystyle{IEEEbib}
\bibliography{refs}

\end{document}